\documentstyle[11pt,aaspp4]{article}

\slugcomment{Accepted for Ap.J.}
\lefthead{Harris et al.}
\righthead{A Very Cool White Dwarf}

\begin{document}

\title{A Very Low-Luminosity, Very Cool, DC White Dwarf\footnote{
Based partly on observations made at the Keck Observatory, the
Steward Observatory, University of Arizona, and Kitt Peak National
Observatory, National Optical Astronomical Observatories.
NOAO is operated by the Association of Universities for
Research in Astronomy, Inc., under cooperative agreement with
the National Science Foundation.
The W. M. Keck Observatory is operated by a scientific partnership
among the California Institute of Technology, the University of
California, and the National Aeronautics and Space Administration.
The Observatory was made possible by the generous financial support
of the W. M. Keck Foundation. }}

\author{Hugh C. Harris, Conard C. Dahn, Frederick J. Vrba, and
Arne A. Henden\footnote{Universities Space Research Association}}

\affil{U.S. Naval Observatory, PO Box 1149, Flagstaff, AZ 86002}

\author{James Liebert and Gary D. Schmidt}

\affil{Steward Observatory, University of Arizona,
933 North Cherry Ave., Tucson, AZ 85721}

\author{I. Neill Reid}

\affil{Palomar Observatory, California Institute of Technology, MS 105-24,
Pasadena, CA 91125}

\begin{abstract}
The star LHS~3250 is found to be a white dwarf at a distance of 30~pc.
Its absolute magnitudes ($M_V = 15.72$; $M_{\rm bol} = 16.2$)
put it among the least-luminous white dwarfs known.
Its optical spectrum shows no features, indicating it has
a DC classification, and it shows no detectable polarization,
indicating it does not have a very strong magnetic field.
However, its broadband colors show it to have
a unique spectral energy distribution, and it stands out from
all other stars in $BVI$ and other broadband photometric surveys.

We discuss these properties, and conclude that LHS~3250 must be
an extremely cool white dwarf with strong collision-induced
absorption at red-infrared wavelengths from molecular hydrogen,
in accord with models for very cool white dwarf atmospheres.
If so, it is the first such star known, and the first star to
provide observational evidence supporting these models.
It suggests that other very cool white dwarfs,
both halo white dwarfs and the oldest disk white dwarfs,
also may have colors affected by similar absorption.
The atmospheric composition of LHS~3250 is not known, and therefore
its temperature is poorly determined.  It may be
a helium-core star with a mass $0.3 < M < 0.45 M_{\odot}$
and a product of mass-transfer in a close binary system.
However, until its temperature is better known,
its mass and age remain uncertain.
\end{abstract}


\section{Introduction}

The star LHS~3250 (LP70-238; WD1653+630) is a faint star discovered
by Luyten (1970; 1976) to have a proper motion of 0.512 arcsec yr$^{-1}$
in a position angle of 275.2${\arcdeg}$
and a color-class ``f''.  These properties indicate that
it is likely to be a white dwarf, and it was observed spectroscopically
by Hintzen (1986) who verified it as being a white dwarf
with a classification of DC9.  USNO parallax observations began
in 1992, and they have now confirmed that the star has the very
low luminosity of a cool white dwarf (discussed in Section 2).
However, in the process of measuring its colors accurately,
we have found it to have a unique spectral energy distribution.
Optical spectra and polarization data have been obtained,
and lead us to conclude that most likely it has an extremely
low temperature.  This conclusion is of particular interest
because low-temperature, low-luminosity white dwarfs in the
Galactic disk can constrain the early star-formation history
of the disk, and because such stars in the Galactic halo
potentially can contribute significant mass to the halo.
In this paper, we discuss the unusual properties of this star,
and attempt to arrive at a plausible description of its nature.

\section{Observational Constraints}

\subsection{Astrometry}

LHS~3250 was observed with the 1.5~m
Strand Telescope and a Tektronix 2048x2048 CCD camera as part of
the USNO parallax program during 1992--1998.  This configuration
gives typical errors of $\pm$0.5 mas in the relative parallax 
with 50 observations or more.  LHS~3250 was observed
with 84 images taken on 81 nights.  The general observing and
reduction procedures for the Tek2048 astrometry are described by
Monet et al. (1992).  The astrometric results are shown in Table 1.
Future observers should note that the coordinates given in the
LHS Catalogue are incorrect by 1.6 arcmin -- 
the LHS Atlas has the correct star marked, and
the correct coordinates are given in Table 1.
Also, the proper motion given in the LHS Catalogue differs
from that given in Table 1 by much more than is normally found
for LHS stars.  We know of no reason for this discrepancy
unless the star was near the faint limit on one of Luyten's plates.

\subsection{Photometry}

Visual $UBVRI$ photometry of LHS~3250 was obtained with the
USNO 1~m telescope using a Tektronix 1024x1024 CCD on four nights
(two in 1995 and two in 1998).
Infrared $JHK$ photometry was obtained with the KPNO 4~m telescope
using the ONIS infrared imager with an ALADDIN InSb detector
on three nights in 1998 for $J$ and $H$ and one night for $K$.
The results are shown in Table 2.  The $RI$ data are on the
Cousins system, and the $JHK$ data are on the CIT system.
Comparison of the data from different nights shows no evidence
for variability larger than about 2\%.

Some of the observed color combinations are displayed in
Figure 1, where the colors of LHS~3250 are compared to the
sample of cool white dwarfs published by Bergeron et al. (1997)
and Leggett et al. (1998).  The $U-B$ colors
for this sample of stars were taken from McCook \& Sion (1998).
The colors of LHS~3250 contrast strongly with other white dwarfs,
providing the first evidence for the unusual properties of this star.
(LHS~1126 is also marked in the figure because it may be a similar
type of star, as we discuss in Section 3.)
The $U-B$ and $B-V$ colors are both consistent with a cool white dwarf
with $T_{\rm eff} \sim 5000$~K, while the $V-I$ and $V-K$ colors
would normally indicate a much hotter temperature.
The flux distribution in Table 2 is strongly peaked near 6000~\AA.
To the eye, this star would appear distinctly orange.

The colors shown in Figure 1 cannot arise from a composite
spectrum, as could be produced by a physical binary system
or a chance alignment with a field star.
Therefore, this region of the color-color plots in Figure 1
is usually empty.  In fact, no other objects are known to fall
near LHS~3250 in Figure 1b other than QSOs with redshift near 3.5-4.0
(\cite{haw96}).  Therefore, stars like LHS~3250 should be identified
easily in a broadband survey using $BVI$ filters or the Sloan
$g^{\prime} r^{\prime} i^{\prime}$ filters.

\subsection{Spectroscopy and Spectropolarimetry}

Low-resolution spectra have been obtained with the Steward Observatory
2.3~m Bok telescope, the Palomar 5~m telescope,
and the Keck II 10~m telescope.
All data show a continuous spectrum with no detectable features,
confirming the DC classification made by Hintzen (1986).
The spectrum with the highest signal/noise is that from
the Keck telescope, a 15-minute exposure taken with the
Low-Resolution Imaging Spectrograph (\cite{oke95}),
with a resolution of 9~\AA.  It is displayed in Figure 2.
(This spectrum was converted from counts to a flux scale
using the other three flux-calibrated spectra and the
broad-band photometry.  Finally, 3-pixel smoothing was applied.)
Beside the wavelength coverage shown in this spectrum, the spectrum
from the Steward 2.3~m telescope with 60 min integration includes
3600-4000~\AA\  (albeit with lower signal/noise).  It is included
in Figure 2, and shows no detectable calcium features.
We also note that Hintzen's spectra covered only the blue spectral
region (3600-5400~\AA) and that he mentioned no calcium features.

The spectral energy distribution shown in Figure 2 is unlike
any other known white dwarf.
A comparison with other spectra of DC and other cool white dwarfs
(e.g., Wesemael et al. 1993, Figure 15; Bergeron et al. 1997, Figure 18)
shows that the sudden decline in flux redward of 6000~\AA\ is
contrary to the normal behavior of cool white dwarfs,
where $F_{\nu}$ reaches a maximum redward of $1 \mu$.

White dwarfs with unusual colors are often found to have both 
distorted spectra and significant circular polarization due
to photospheric emission in a strong magnetic field ($B>10^8$ G).
To investigate this possibility, 
data were also obtained with the CCD Spectropolarimeter
(\cite{sch92}) on the Steward Observatory 2.3~m Bok 
telescope on two occasions in 1998.  In Figure 3 are displayed 
polarization spectra as $V/I (\%)$ resulting from sequences
totaling 40 min and 53 min on 1 Apr. and 23 Sep., respectively.
Both occasions show a distinct lack of significant circular
polarization on either side of the 6000\AA\ spectral break, 
and the 3$\sigma$ upper limit
for the spectrum-added polarization is $V/I < 0.4\%$ in each dataset.
We conclude that magnetism is a very unlikely explanation for
the peculiar energy distribution of LHS 3250.

\subsection{Color-Magnitude Diagram}

With the absolute magnitude (Table 1) and observed colors (Table 2),
we can construct color-magnitude diagrams that are usually indicative
of the approximate temperature and mass of white dwarfs.
Figure 4 shows the $M_V$ vs. $B-V$ diagram in panel (a) and
$M_V$ vs. $V-I$ in panel (b).
In this figure, LHS~3250 is compared with other
white dwarfs with accurate parallaxes obtained with CCDs at USNO,
plus the unusual low-luminosity star ESO439-26 (\cite{rui95}).
The points include all white dwarfs from Monet et al.\ (1992)
plus other white dwarfs with parallaxes determined since 1992
but not yet published;  all points are shown with their formal
one-sigma error bars.
The curves show the model-atmosphere colors of white dwarfs with
pure hydrogen and pure helium atmospheres (\cite{ber95b}).
The four white dwarfs known to have high surface gravity and
high mass from spectrophotometric studies are labelled; they stand
out from other stars in both panels.  From this figure, one would
derive for LHS~3250 a somewhat high mass (near 0.8 $M_{\odot}$) and a
cool temperature from the $B-V$ color, but an extraordinarily-high
mass (near 1.4 $M_{\odot}$) and high temperature from the $V-I$ color.
This discrepancy is another symptom of the peculiar colors of
LHS~3250 shown in Figure 1.  In the following discussion,
we conclude that probably neither diagram gives a correct
understanding of the star because its colors are far from
being indicative of its temperature or gravity.

\subsection{Luminosity}

In order to establish the luminosity of LHS~3250,
the bolometric correction (BC) is needed.  Using
\begin{displaymath}
{\rm BC = BC_{\sun}
 + 2.5~log \left( \frac{\int {\it F}_{\lambda_{\star}}
      {\it S_{V}} ( \lambda ) d \lambda}
   {\int {\it F}_{\lambda_{\star}} d \lambda} \right)
 - 2.5~log \left( \frac{\int {\it F}_{\lambda_{\odot}}
      {\it S_{V}} ( \lambda ) d \lambda}
   {\int {\it F}_{\lambda_{\odot}} d \lambda} \right) }
\end{displaymath}
where $S_{V}({\lambda})$ is the transmission function of the $V$ filter,
we derive BC~$ = 0.45 \pm 0.10$.  The uncertainty is due
to the coarse sampling of the spectrum shown in Figure 5.
This value is quite different from that for normal cool white dwarfs
which have BC~$ = -0.2$ at $T_{\rm eff} = 5000$~K to
BC~$ = -1.0$ at $T_{\rm eff} = 4000$~K (\cite{ber95b}).
However, positive values of BC are reproduced in the
models of very cool white dwarfs with strong collision-induced
absorbtion (CIA; see Section 3.2)
owing to their large infrared flux deficiency.
We then derive $M_{\rm bol} = M_{V} + {\rm BC} = 16.17 \pm 0.11$
and log($L/L_{\sun}) = -4.57 \pm 0.04$.
With the exception of the high-mass star ESO439-26,
this is the lowest luminosity known for any white dwarf.

\subsection{Stellar Population}

The low luminosity of LHS~3250 raises the question of its
membership in the Galactic disk or halo.
Can we otherwise determine its origin, and perhaps thereby
place some constraints on its age? Its tangential
velocity (Table 1) is typical of a white dwarf in the old disk.
However, its Galactic coordinates ($l = 94\arcdeg$, $b = 38\arcdeg$)
mean that its proper motion reflects primarily its $U$ and $W$
space velocities, while its $V$ velocity is mostly projected into
the unknown radial velocity.  One could hypothesize a radial velocity
in the range --170 to  --220~km~s$^{-1}$ which would result in
$V$ velocity components in the range --180 to --220~km~s$^{-1}$,
respectively, and $W$ velocity components in the range
--50 to --80~km~s$^{-1}$, respectively, values typical of stars
in the Galactic halo.  However, the DC nature of the spectrum
offers no possiblity for testing this hypothesis.
Therefore, its kinematics do not identify its population of origin.

\section{Temperature Constraints}

Despite the fact that the luminosity of LHS~3250 is well determined,
its effective temperature must be defined more accurately before
we can proceed further.  Everything else about the star
(the mass, the radius, the cooling age, and the evolutionary state)
depends on determining $T_{\rm eff}$ accurately.
In this section, we attempt to constrain the temperature from the
observed spectral energy distribution.  Then in the following section
(Section 4), we discuss further constraints imposed from the accurate
luminosity and other considerations.

The unique spectral energy distribution is given in Table 2
and shown in Figure 5.
A comparison with white dwarf model colors (\cite{ber95b})
shows that a model star with $T_{\rm eff} \sim 5000$~K
matches the $UBVR$ data for LHS~3250 quite well,
while a model star with $T_{\rm eff} \sim 15000$~K
and a helium-dominated atmosphere matches the $RIJHK$ data.
Apparently strong absorption is occuring in the atmosphere
of LHS~3250, either at blue wavelengths if the star is hot,
or at red/near-infrared wavelengths if it has a cool temperature.
First we consider the possibility that it is hot.

\subsection{A High Temperature?}

Hot white dwarfs normally have hydrogen or helium lines
visible:  DA stars with hydrogen atmospheres have H${\alpha}$
present with an equivalent width of $>$1~\AA\ when the temperature
is $>5300$~K (\cite{ber97}), while DB stars with helium
atmospheres have HeI lines present at ${\lambda\lambda}$4471, 5876,
and 6678 when the temperature is $>11000$~K (\cite{wes93}).  
In contrast, DC stars are known with classifications as hot as
class DC5, with $B-V$ colors $\sim$0.00, indicating temperatures
$\sim 12000$~K for helium-dominated atmospheres;
at hotter temperatures helium lines become visible.
Because LHS~3250 shows neither hydrogen nor helium lines,
its temperature must be $<12000$~K,
but this temperature is not hot enough to match the $RIJHK$ data.

The hot-star alternative also requires some {\it continuous}
absorption at blue wavelengths to explain the energy distribution in
Figure 5.  However, possible sources of continuous absorption (molecules,
for example), will not be present in a high-temperature atmosphere.
The DZ (metallic-lined) and DX (peculiar) white dwarfs sometimes
have broad absorption features [e.g., G99-44, LP701-29, and G240-72
(\cite{wes93})], but only when they are cool; hot white dwarfs never
show such broad absorption so as to be classified DC.
Most of the stars with broad (often unidentified) features
also have identified metallic features, primarily Ca lines,
which are not seen in LHS~3250 (Section 2).

A further consequence of a high temperature for LHS~3250 would
be that it would have a high mass, probably near the Chandrasekhar
limit, as shown in Figure 4(b).  This would be required by its faint
absolute magnitude, which places it 2-3 mag fainter than any white
dwarfs of similarly high temperature.
ESO439-26, for example (see Figure 4), has a temperature near
4500~K and a mass ${\sim}1.2 M_{\odot}$ (\cite{rui95}).
A high mass (and resulting high gravity)
is possible, but it probably would not help explain the peculiar colors,
as gravity has virtually no effect on the colors of hot,
helium-dominated atmospheres (\cite{ber97}).  Therefore,
we believe that a high temperature for LHS~3250 is unlikely.

\subsection{A Cool Temperature?}

Can the unusual energy distribution be explained with a cool
temperature?  Molecular opacities are potentially important in
the high gravity of a cool white dwarf atmosphere and might supply
the continuous opacity apparently needed to understand LHS~3250.
A possible explanation is that collision-induced absorption (CIA)
by molecular hydrogen\footnote{
Collisions of H$_2$ molecules with other H and He atoms and molecules
become important in a high-density atmosphere.  These collisions
induce a temporary dipole moment in the H$_2$ molecule that creates
permitted dipole transitions and a greatly increased opacity
at red and near-infrared wavelengths (\cite{lin69}; \cite{len91};
\cite{sau94}).  The effect is greatest in a helium-dominated
atmosphere that is otherwise very transparent.
}
is dominating the opacity for wavelengths $>$6000~\AA. 

The effects of CIA on the colors of cool white dwarfs with various
hydrogen/helium compositions have been described by Bergeron et al. (1995a)
and Hansen (1999).  Various models for white dwarfs with extremely
low temperatures and pure hydrogen atmospheres (\cite{ber97};
\cite{han98a}; \cite{han98b}; \cite{han99}; \cite{sau99}; \cite{cha99})
agree in predicting strong flux-deficiencies for $\lambda > 1\mu$,
a peak in the flux distribution near $\lambda \sim 6000-9000$~\AA,
and blue $V-I$ and $V-K$ colors.  However, high-density helium
atmospheres are difficult to model, and no models of mixed-composition
atmospheres are presently available for temperatures below 4000~K.
The models that are available predict stronger infrared
absorption, and a bluer wavelength at which the flux reaches a maximum,
for lower temperatures and for He/H ratios of roughly 10 to 100.
The blue $V-K$ color of LHS~1126 (a star noted in Figure 1c) was
explained by this mechanism (\cite{ber94}; \cite{sch95}; \cite{ber97}).
Here, we see that LHS~3250 has much more extreme colors
than LHS~1126, so the effect of CIA must be much stronger
if it is the explanation for the colors of LHS~3250.

A start at fitting the spectral energy distribution of LHS~3250 has been
made by Bergeron (1999), and the result kindly made available to us.
The composition was assumed to be pure hydrogen.  A grid of model
atmospheres extending from 4000K to 1500K was calculated at intervals
of 250K for log~$g =$~7.0, 7.5, 8.0, 8.5, 9.0.  The models
were calculated as described in Bergeron et~al. (1995a).
The equation of state was modified to include molecular H$_2$, H$_2^+$
H$_3^+$ as in Saumon et~al. (1994),
utilizing the opacity calculations of Lenzuini et~al. (1991) with the
CIA coefficients for H$_2$, H and He species from Borosow et~al. (1989).
Similar models were used previously for detailed analyses of cool
white dwarfs (\cite{ber97}; \cite{leg98}).  Due to the lack of
evolutionary models (such as the C/O core grid of Wood used in
earlier papers), zero temperature results for helium cores were adopted
(\cite{ham61}), and should provide an adequate mass-radius relation
at these low temperatures.

An approximate fit to the $VRIJHK$ flux distribution,
using a pure hydrogen atmosphere, is shown in Figure 5 (middle panel).
The result is an extremely low temperature ($T_{\rm eff} = 2020$~K),
where the fit has been done assuming a plausible value of log~$g = 6.8$.
The model is qualitatively close to the data, and shows that strong CIA
in a hydrogen atmosphere with very low temperature is possible.
A similar temperature near 2100~K would be derived from
the models with pure hydrogen by Saumon \& Jacobson (1999),
and presumably with those by Chabrier (1999), but temperatures are
not included for the latter models.

However, we discuss below some difficulties that arise with a
temperature this low and a pure hydrogen atmosphere.
Instead, helium-dominated atmosphere models (\cite{ber95a})
may be more realistic.  In these models, CIA 
is strong at temperatures as high as 4000~K, and becomes
strongest (at a given temperature) for He/H compositions of 10-100.
One such model is also shown in Figure 5 (bottom panel).
It {\it does not} reproduce the flux distribution of LHS~3250 in detail,
but it does reproduce the strong infrared flux deficiency seen.
Apparently a temperature lower than 4000~K is required
to give increased CIA opacity to match the data.
Therefore, we conclude that the temperature is in the range
$2000 \leq T{\rm_{eff}} < 4000$~K,
and the He/H ratio is near zero if the temperature is at the low end
of the range, and is large (perhaps 10-100) if the temperature
is at the high end of the range.

\section{Discussion and Conclusions}

In this section, we explore the physical parameters of LHS~3250
(the radius, mass, and cooling age) and its evolutionary state.
These quantities are directly related to its well-determined
luminosity and its approximately-known temperature.
A detailed analysis will require mixed-composition model atmospheres
at cooler temperatures than are available now, and is beyond the
scope of this paper.  Our present goal is to constrain
the physical parameters (if possible) by considering a range
of possible temperatures ($2000 \leq T{\rm_{eff}} < 4000$~K, as
determined above), and exploring the consequences of each temperature.
Table 3 gives a summary of these results.  Here, the second
column lists the radius derived for each possible temperature,
using the measured luminosity above.  This is followed by the mass,
as derived from mass-radius relations for white dwarfs with different
core and atmospheric compositions.  For this purpose, we give the mass
for each value of the radius found using different models listed in the
footnotes to Table 3.  Finally, the table gives the (very approximate)
cooling time to reach $L/L_{\odot} = -4.6$, taken from the same models.
Table 3 shows that (1) the mass depends strongly on the
adopted temperature, (2) the exact mass depends on the presence
and thickness of a hydrogen atmosphere, and 
(3) a helium core and a mass lower than normal for a white dwarf
are very possible, but a normal mass and a carbon/oxygen core are
also possible.

Because a helium core and a reduced mass must be a result of
mass transfer in a binary system, and because such mass transfer
often leaves a double white dwarf system with the two stars having
nearly equal masses, we have added a second set of values to Table 3
for the case of a composite of two equal-luminosity white dwarfs.
For such a composite system, the luminosity of each star
at a given possible temperature
is less than for the single-star case by a factor of 2,
the radius is less by a factor of $\sqrt{2}$, the mass is greater
by a factor of $\sim$1.5-2.5, and the cooling age to reach
$L/L_{\odot} = -4.9$ is generally much greater.

Table 3 indicates that LHS~3250 could have a wide variety of states.
First, if LHS~3250 has a C/O core, it could be a single star
with a mass near 0.50-0.55~$M_{\odot}$, a temperature
near 3800~K, and a cooling time of about 9~Gyr.
A helium-dominated atmosphere would be required to reproduce the
observed spectral energy distribution at this relatively
high temperature (Section 3).
Second, if LHS~3250 has a helium core with a helium-dominated
atmosphere, Table 3 shows that it could have a wide range of
temperatures ($2000 < T{\rm_{eff}} < 3500$~K)
and masses ($M < 0.45 M_{\odot}$).
It could be a single visible star or a composite double white dwarf
with somewhat lower temperatures and masses.
Finally, if LHS~3250 has a helium core and a hydrogen atmosphere,
then the spectral energy distribution requires a temperature
near 2000~K.  Then Table 3 shows that a very low mass, $<0.15 M_{\odot}$,
is required to give the large radius demanded by such a low temperature.

Evolutionary models for He-core stars by Driebe et al. (1999)
may further limit the states that are possible for LHS~3250.
These models are intended
to include a realistic thickness of hydrogen envelope on low-mass
stars based on when and how fast mass is lost during evolution
on the red giant branch.  These authors argue that a thick hydrogen
envelope will inevitably survive on He-core stars with
$M < 0.3 M_{\odot}$, and that the structure of the
stars will give prolonged H-burning, resulting in the cooling times
being extended significantly.
If these evolutionary models are correct, then a mass
${\leq}0.25 M_{\odot}$ would result in a cooling time greater than
the age of the universe to reach the low luminosity of LHS~3250,
and must be ruled out.  This conclusion eliminates all states with
very low-mass, including all states with a hydrogen atmosphere.
Only two remaining states are allowed:
a single He-core white dwarf with a
temperature in the range $3200 < T{\rm_{eff}} < 3500$~K,
a mass 0.3-0.45~$M_{\odot}$, and a cooling time of roughly 12~Gyr;
or a composite of two He-core white dwarfs with a temperature
close to 2700~K, a mass ${\sim}0.3 M_{\odot}$,
and a cooling time ${\sim}15$~Gyr.  Only if somehow the
predicted thick H envelope has been lost (or perhaps some H has
been accreted since the white dwarf formed with a helium atmosphere),
then perhaps a thin H atmosphere, with the very low temperature
and low mass, is still possible.  Both ``thin atmosphere'' models
(\cite{han98a}; \cite{ben98}; \cite{ben99}) predict that convection
will mix helium up into the atmosphere, diluting the hydrogen
as a low-mass star cools.  However, below roughly 4000~K,
a radiative zone develops (\cite{han98a}) and diffusion should
remove the helium, again leaving a pure-H atmosphere.
This scenario appears to be the only possibility that LHS~3250
has a hydrogen-dominated atmosphere.

Applying these constraints, we can place LHS~3250 in the H-R diagram.
Figure 6 shows LHS~3250, again compared with the same sample of
cool white dwarfs shown in Figure 1 (\cite{ber97}; \cite{leg98}).
LHS~3250 has a lower luminosity than any star except the massive
white dwarf ESO439-26 (\cite{rui95}).
LHS~3250 is plotted with the possible temperatures derived above.

A He-core white dwarf must have lost mass during its
giant-branch evolution through mass transfer in a close binary system.
Also, the composite double white dwarf states discussed above
might have the two stars comprising a close binary system,
and possibly having exchanged mass.  Therefore, LHS~3250 may be
(or may once have been) a binary.  
The companion could be a neutron star, and thus not detectable
unless it was also a pulsar; then LHS~3250 would be like an older
version of the low-mass white dwarfs found to be the visible
companions of some millisecond pulsars.
Alternatively, we may be seeing a composite of two white dwarfs,
both with similar low temperatures.
Close binaries which undergo common-envelope evolution
are believed to be the likely progenitors of double-degenerate
stars with a He-core component (e.g. \cite{ibe93}; \cite{ibe97}).
These progenitors can, and apparently often do, produce a pair
of nearly identical stars.
Examples are the pair of C/O white dwarfs L870-2 (\cite{saf88}),
and the pairs of He-core white dwarfs L101-26 (\cite{mor97})
and Ton1323 (\cite{mar95}).
Of the other known close double-degenerate stars, most have
a He-core primary star (\cite{saf98}), with a secondary star
of unknown type.  Therefore, arriving at a close binary system
with both components being such very cool (and very old) He-core
stars is plausible.

Stars related to LHS~3250 have recently been found in the
globular cluster NGC~6397 (\cite{coo99}; \cite{edm99}).
Three ``non-flickering'' stars appear to be He-core white dwarfs
with masses $<0.4 M_{\odot}$, and the one with an observed
spectrum has a hydrogen atmosphere with $T_{\rm eff} \sim
17000$~K and a mass ${\sim} 0.25 M_{\odot}$.  All three stars
are much hotter and more luminous than LHS~3250, and are much
younger (more recently formed) white dwarfs.
Thus the three may eventually evolve to be similar to LHS~3250.

In conclusion, we find LHS~3250 to have a temperature
sufficiently low that its colors are strongly affected by some
continuous absorption, probably CIA from molecular hydrogen.
As such, it provides the first example of a very cool white dwarf
with blue $RIJHK$ colors, and it supports models of white dwarfs
with H or H/He atmospheres
(\cite{ber95a}; (\cite{ber97}; \cite{han98a};
\cite{han98b}; \cite{han99}; \cite{sau99}; \cite{cha99})
that show their $RIJHK$ colors becoming bluer with decreasing
temperature and luminosity.  Finding other white dwarfs
with low (and lower) luminosities and temperatures
is currently of interest for determining the density and
luminosity function of old disk and halo white dwarfs.
The population (disk or halo) to which LHS~3250 belongs
is not known.  It may be a product of mass
transfer in a close binary system, not a result of single-star
evolution.  Nevertheless, its noteworthy colors at this
very low temperature support the conjecture that many
low-luminosity white dwarfs in the halo will have blue colors.
Stars like LHS~3250 will be easy to identify in photometric
surveys, so we should soon measure how frequently they occur.

\acknowledgments

We thank P. Bergeron for calculating low-temperature models of
pure hydrogen white dwarfs and making available his fit to our data.
The astrometric data were acquired as part of the Naval Observatory
parallax program, including observing by H. Guetter, C. Luginbuhl,
A. Monet, D. Monet, J. Pier, R. Stone, and R. Walker.
This work has been supported by the National Science Foundation
through grant 97-30792 to G.S. and grant 92-17961 to J.L.

\clearpage


\figcaption[harris_fig1.ps]{Color-color diagrams showing LHS~3250
is easily distinguished from a sample of cool white dwarfs.
\label{fig1}}

\figcaption[harris_fig2.ps]{Spectrum of LHS~3250 taken with the
Keck II telescope, plus the blue end of a spectrum taken with the
Steward 2.3~m telescope (shifted vertically by 10 flux units for
clarity) in order to show wavelengths blueward of 4000 \AA.
\label{fig2}}

\figcaption[harris_fig3.ps]{Spectropolarimetry of LHS~3250 taken with
the Steward 2.3~m telescope.  The data for 23 Sep. have been displaced
upward by 20\% for clarity.
\label{fig3}}

\figcaption[harris_fig4.ps]{Color-absolute magnitude diagrams of LHS~3250
compared to white dwarfs with accurate CCD parallaxes.  Curves show
model atmosphere colors (\cite{ber95b}) for pure hydrogen atmospheres
(solid curves) and pure helium atmospheres (dashed curves).
Four stars known to have high mass are labelled (see text).
Note the discrepant position of LHS~3250 between the two panels.
\label{fig4}}

\figcaption[harris_fig5.ps]{Spectral energy distribution of LHS~3250.
The top panel gives the observed distribution, with error bars.
The middle and bottom panels show the observed data compared with
two models having strong CIA opacity.
\label{fig5}}

\figcaption[harris_fig6.ps]{H-R diagram showing LHS~3250 with a sample
of cool white dwarfs.  LHS~3250 is plotted with two filled circles
to show a plausible range of (cool) temperatures.
\label{fig6}}




 \clearpage
 \plotone{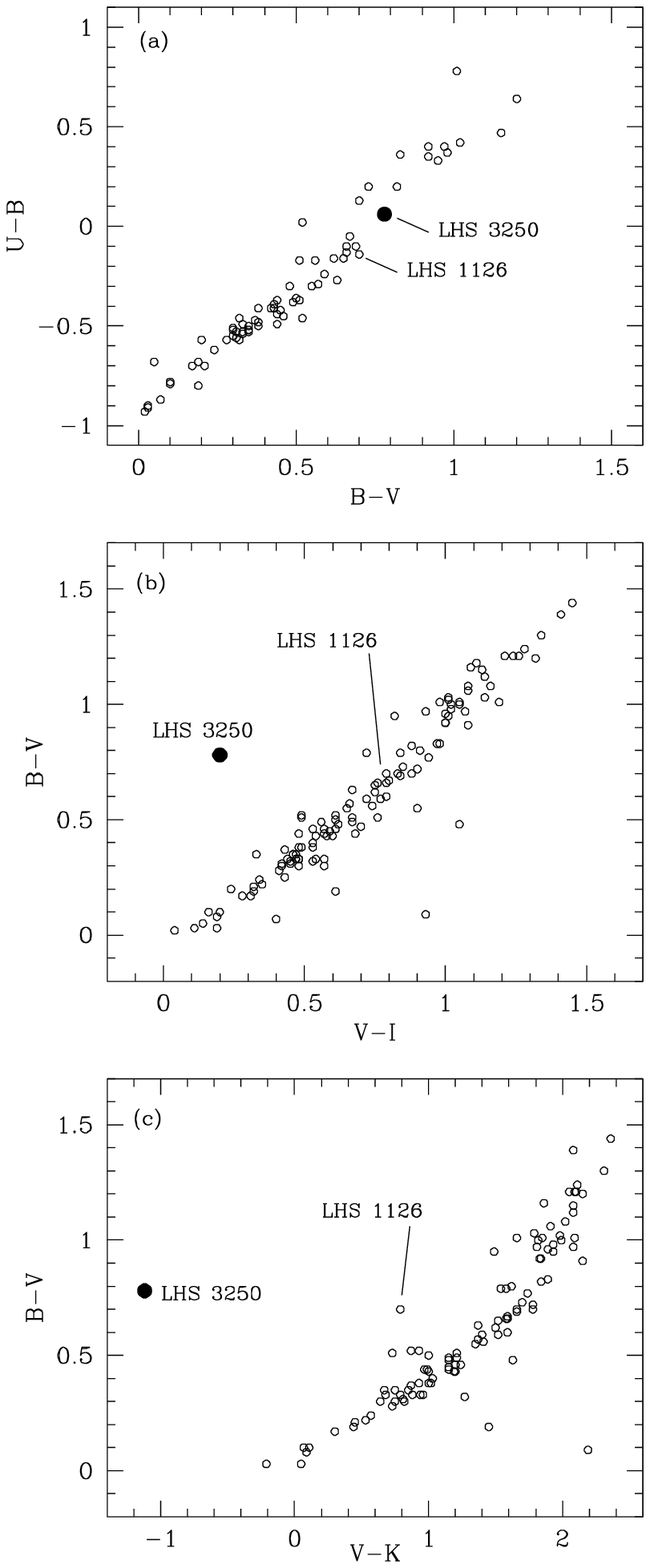}
 \clearpage
 \plotone{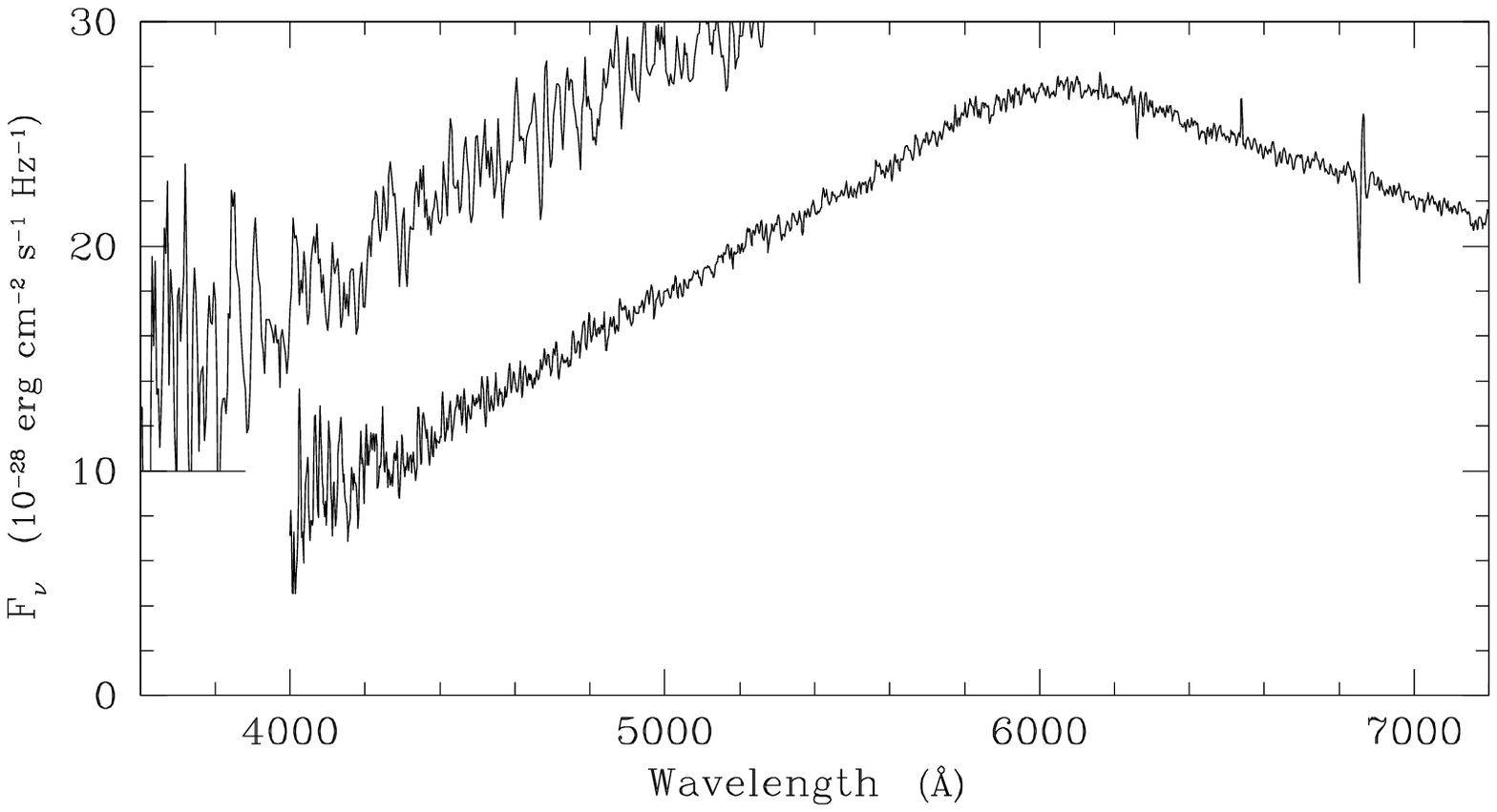}
 \clearpage
 \plotone{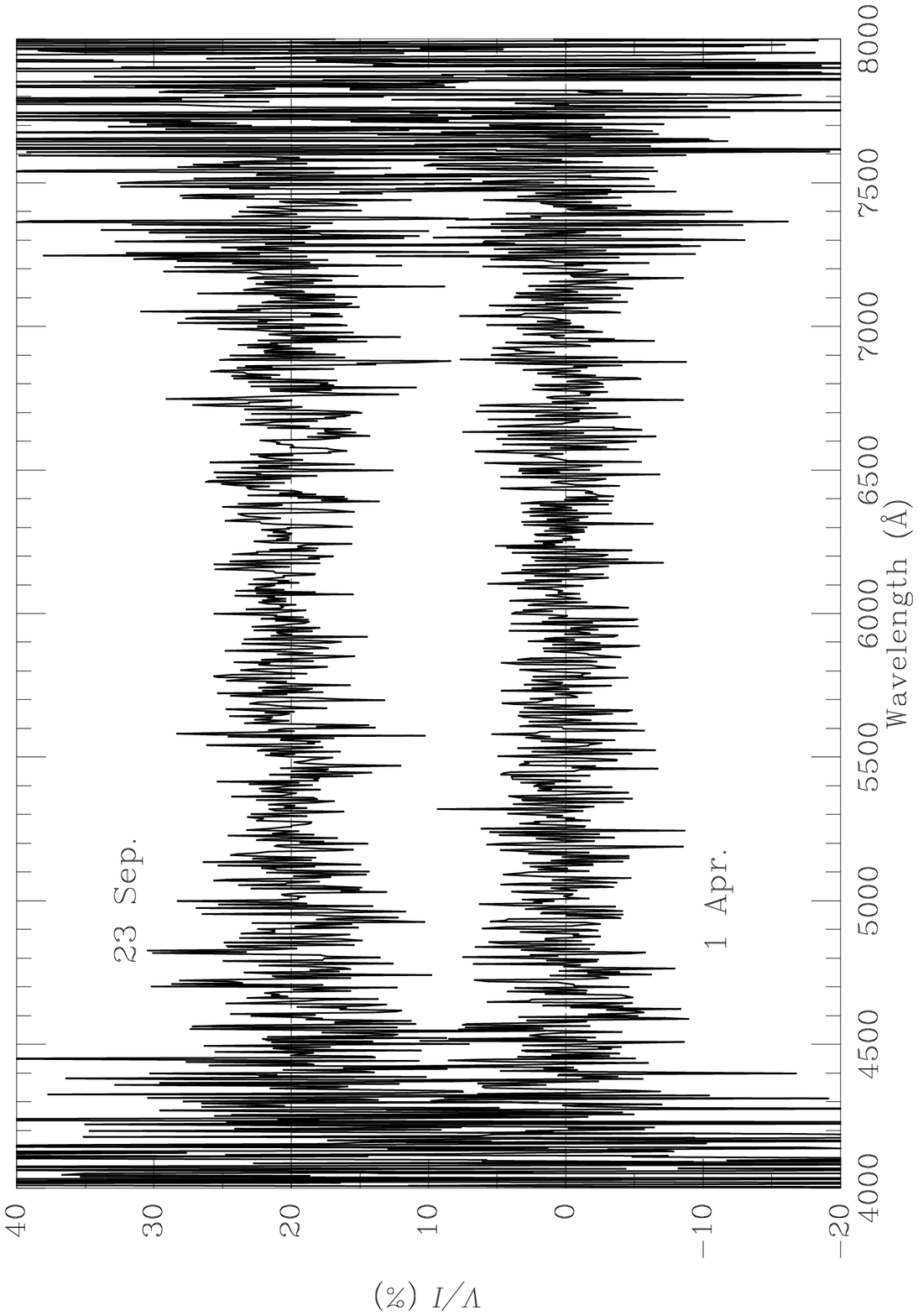}
 \clearpage
 \plotone{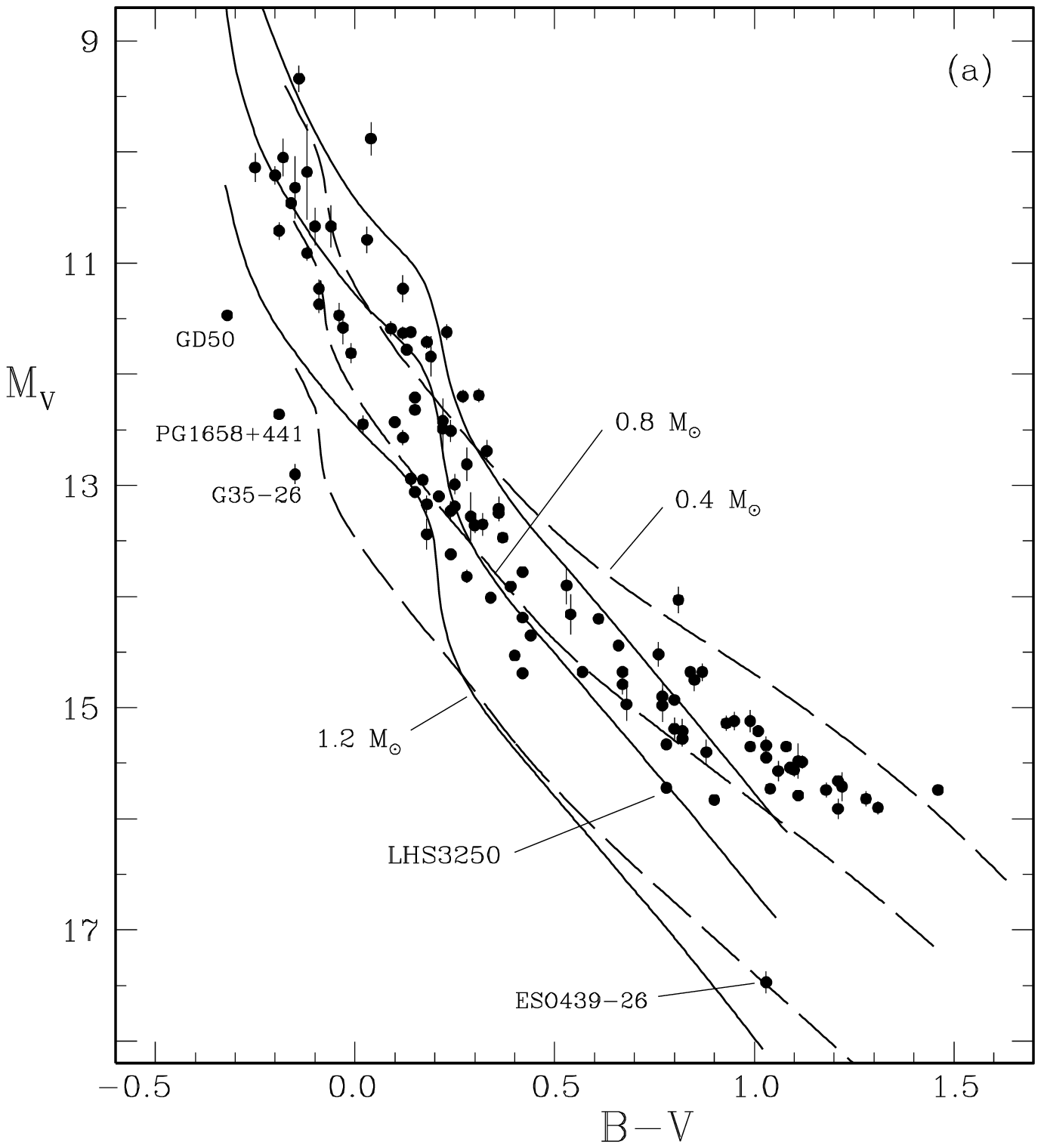}
 \clearpage
 \plotone{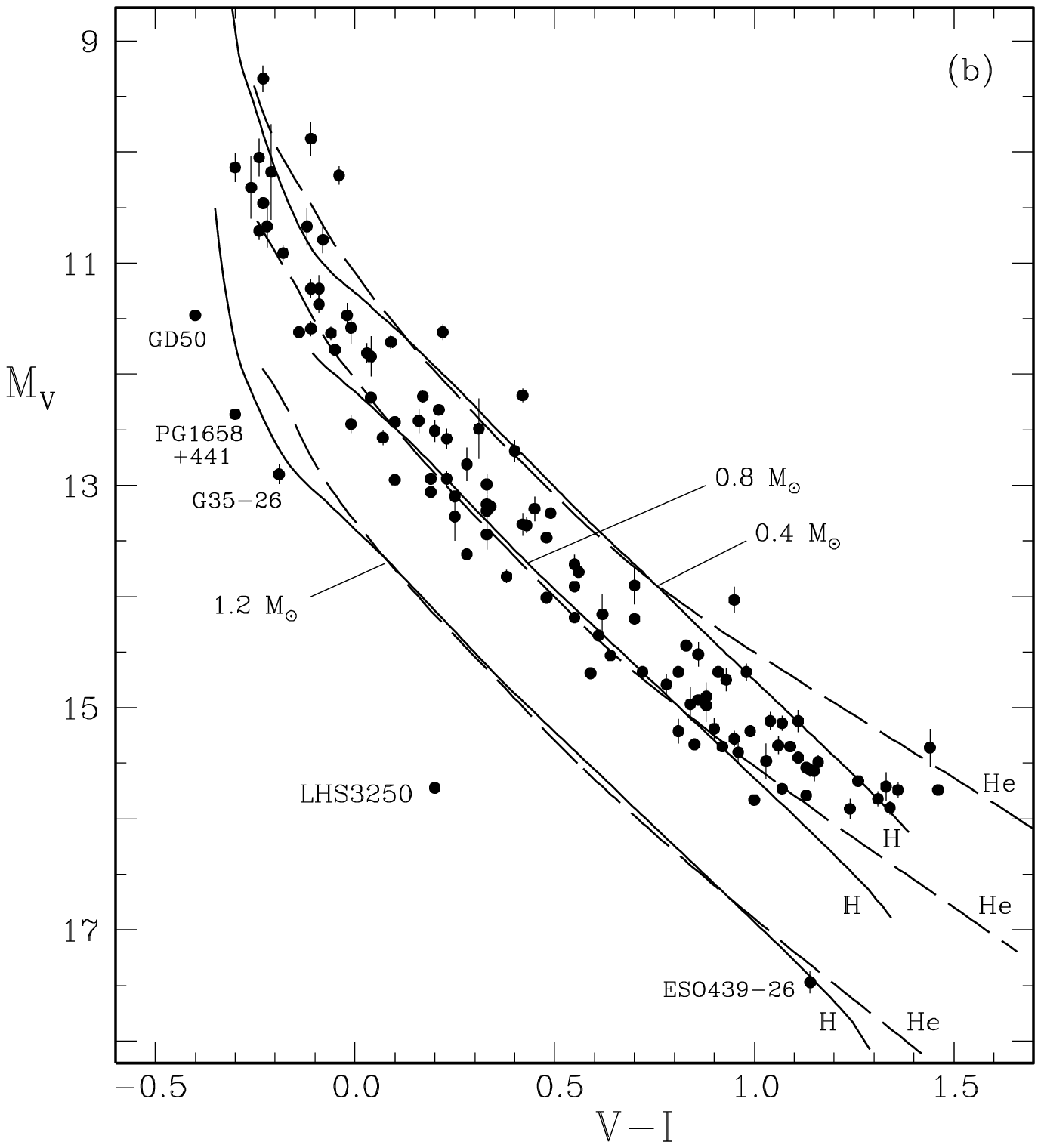}
 \clearpage
 \plotone{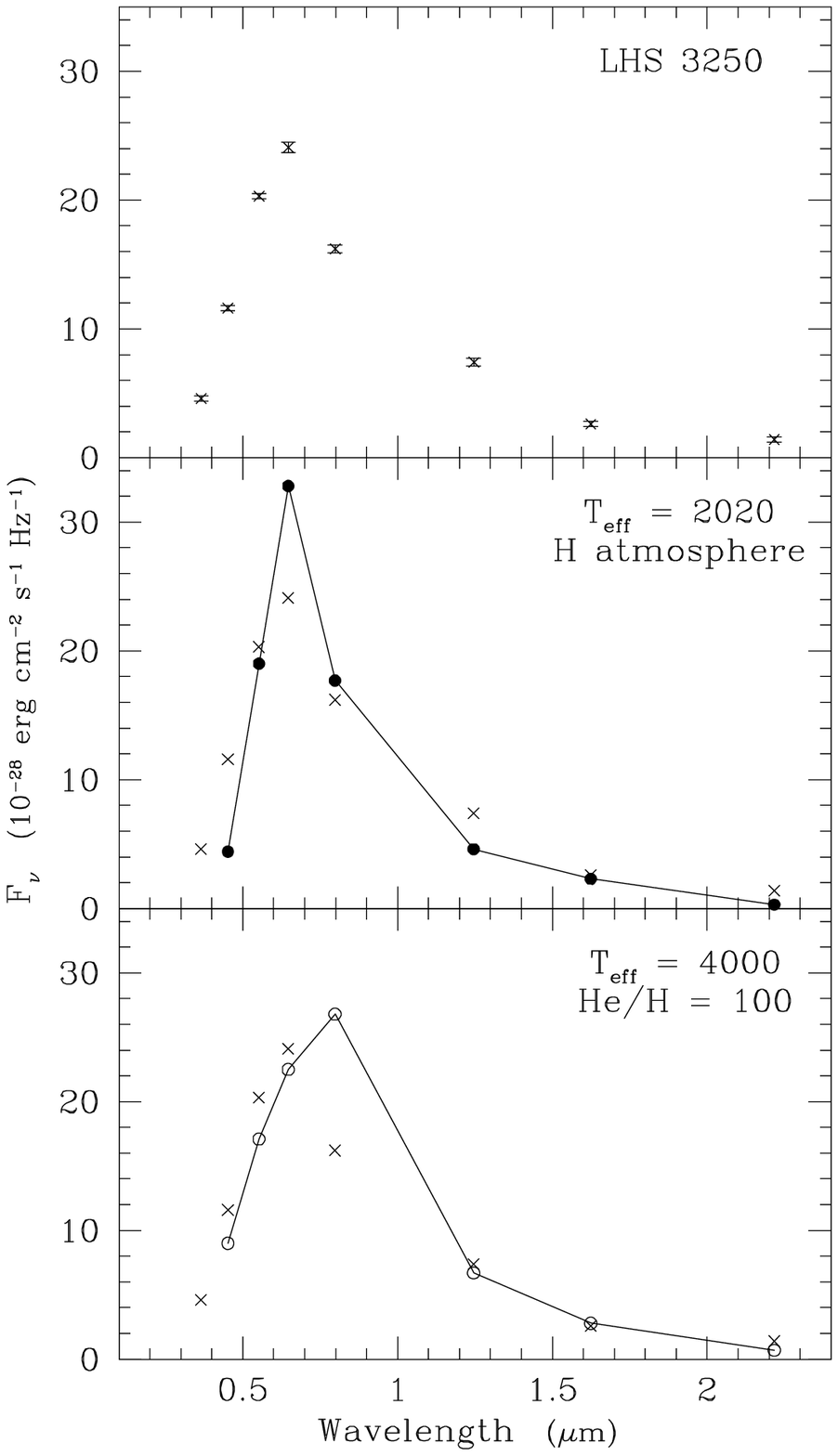}
 \clearpage
 \plotone{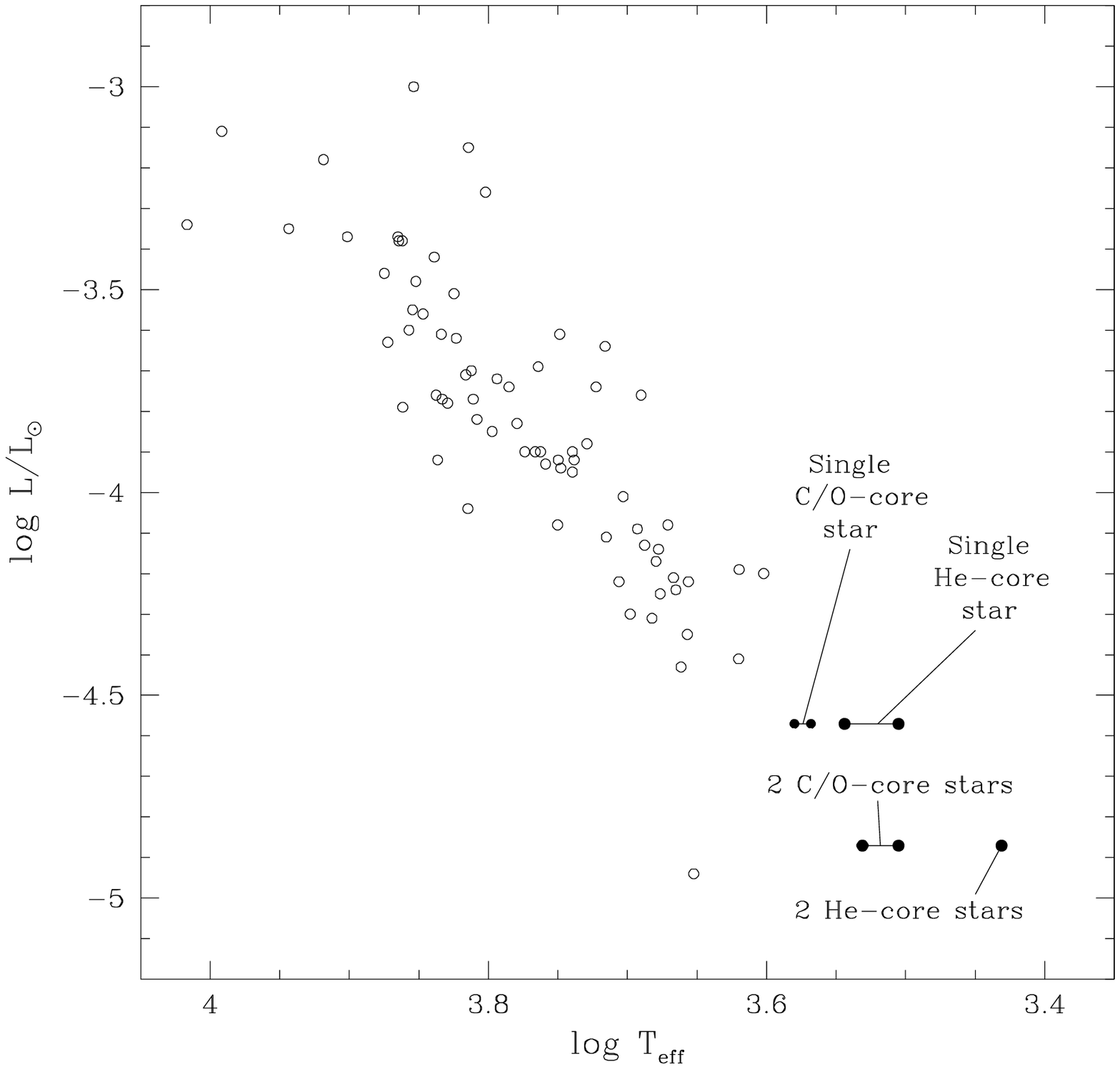}

\clearpage

\begin{deluxetable}{ll}
\tablenum{1}
\tablewidth{2.7in}
\tablecaption{Astrometric Data}
\tablehead{
\colhead{Parameter} &
\colhead{Value}
}
\startdata
RA{\tablenotemark{a}} & $\rm 16{^h}$ $\rm 54{^m}$ $\rm 01.3{^s}$ \nl
Dec{\tablenotemark{a}}& $\rm 62{^o}$ $53{^\prime}$ $\rm 55{^{\prime\prime}}$\nl
$\pi_{\rm rel}$      & 32.26 $\pm$ 0.49 mas \nl
$\Delta\pi$          & 0.78 $\pm$ 0.11 mas \nl
$\pi_{\rm abs}$      & 33.04 $\pm$ 0.50 mas \nl
Distance             & 30.3 $\pm$ 0.5 pc \nl
$\mu_{\rm rel}$      & 565.7 $\pm$ 0.1 mas yr$^{-1}$ \nl
PA                   & 286.11 $\pm$ 0.02 degrees \nl
$V_{\rm tan}$        & 81.2 $\pm$ 1.2 km s$^{-1}$ \nl
$M_V$                & 15.72 $\pm$ 0.04 \nl
\enddata
\tablenotetext{a}{Coordinates are given for equinox and epoch 2000,
derived from the USNO-A Catalog.}
\end{deluxetable}

\clearpage

\begin{deluxetable}{crccr}
\tablenum{2}
\tablewidth{3.5in}
\tablecaption{Photometric Data}
\tablehead{
\multicolumn{2}{c}{Color{\tablenotemark{a}}}  &
\multicolumn{2}{c}{Magnitude}     &
\colhead{$f_{\nu}${\tablenotemark{b}}}
}
\startdata
     &                & $U$& 18.91$\pm$0.05 &  4.6$\pm$0.2 \nl
$U-B$&$ 0.06 \pm$0.05 & $B$& 18.85$\pm$0.02 & 11.6$\pm$0.2 \nl
$B-V$&$ 0.78 \pm$0.02 & $V$& 18.07$\pm$0.01 & 20.3$\pm$0.2 \nl
$V-R$&$ 0.33 \pm$0.02 & $R$& 17.74$\pm$0.02 & 24.1$\pm$0.4 \nl
$R-I$&$-0.13 \pm$0.03 & $I$& 17.87$\pm$0.02 & 16.2$\pm$0.3 \nl
$I-J$&$-0.46 \pm$0.05 & $J$& 18.33$\pm$0.05 &  7.4$\pm$0.3 \nl
$J-H$&$-0.61 \pm$0.07 & $H$& 18.94$\pm$0.06 &  2.6$\pm$0.2 \nl
$H-K$&$-0.25 \pm$0.15 & $K$& 19.19$\pm$0.14 &  1.4$\pm$0.2 \nl
\enddata
\tablenotetext{a}{$RI$ are on Cousins system; $JHK$ are on CIT system.}
\tablenotetext{b}{Units are $10^{-28}$ erg cm$^{-2}$ s$^{-1}$ Hz$^{-1}$.}
\end{deluxetable}

\clearpage

\begin{deluxetable}{ccrlrlrlrl}
\tablenum{3}
\tablewidth{6.4in}
\tablecaption{Possible Radius, Mass, and Cooling Age}
\tablehead{
\colhead{$T_{\rm eff}$} &
\colhead{$R/R_{\odot}$} &
\colhead{$M/M_{\odot}$} & \colhead{Age} &
\colhead{$M/M_{\odot}$} & \colhead{Age} &
\colhead{$M/M_{\odot}$} & \colhead{Age} &
\colhead{$M/M_{\odot}$} & \colhead{Age} \\
\colhead{} & \colhead{} &
\multicolumn{2}{c}{C/O core} &
\multicolumn{2}{c}{He core} &
\multicolumn{2}{c}{He core} &
\multicolumn{2}{c}{He core} \\
\colhead{} & \colhead{} &
\multicolumn{2}{c}{H/He atmosphere\tablenotemark{a}} &
\multicolumn{2}{c}{He atmosphere\tablenotemark{b}} &
\multicolumn{2}{c}{Thin H atmosphere\tablenotemark{c}} &
\multicolumn{2}{c}{Thick H atmosphere\tablenotemark{d}}
}
\startdata
\multicolumn{10}{c}{For single white dwarf:} \nl
4000 &0.0110 &0.69 &{\phn}9 Gyr & & & & & & \nl
3750 &0.0124 &0.52 &{\phn}9 Gyr & & & & & & \nl
3500 &0.0143 &&&{\phn}0.43 &13 Gyr & & & 0.45 & {\phs}12 Gyr \nl
3250 &0.0166 &&&{\phn}0.30 &11 Gyr &0.37&{\phs}14 Gyr&0.34&{\phs}13 Gyr \nl
3000 &0.0194 &&&{\phn}0.20 &{\phn}9 Gyr&0.23&{\phs}10 Gyr&0.24&$>$15 Gyr \nl
2750 &0.0231 &&&{\phn}0.13 &{\phn}7 Gyr&0.14&{\phs\phn}8 Gyr&0.18&$>$15 Gyr\nl
2500 &0.0280 &&&&&0.09&{\phs\phn}6 Gyr&$<$0.15 &$>$15 Gyr \nl
2250 &0.0346 & & & & & & & & \nl
\tableline
\tablevspace{0.1 in}
\multicolumn{10}{c}{For double white dwarf:\tablenotemark{e}} \nl
4000 &0.0078 &0.95 &{\phn}9 Gyr &&&&&& \nl
3750 &0.0088 &0.87 &{\phn}9 Gyr &&&&&& \nl
3500 &0.0101 &0.76 &10 Gyr &&&&&& \nl
3250 &0.0118 &0.59 &11 Gyr &&&&&& \nl
3000 &0.0138 &&& 0.46& 20 Gyr&&& 0.5{\phn} &$\sim$15 Gyr \nl
2750 &0.0164 &&& 0.31& 16 Gyr& 0.38 &$>$15 Gyr & 0.35 &$\sim$15 Gyr \nl
2500 &0.0198 &&& 0.19& 13 Gyr& 0.21 &$\sim$15 Gyr & 0.23 &$>$15 Gyr \nl
2250 &0.0244 &&& 0.11& 10 Gyr& 0.12 &{\phs\phn}9 Gyr& 0.18 &$>$15 Gyr \nl
\enddata
\tablenotetext{a}{Models from Wood (1995) and Hansen (1999).}
\tablenotetext{b}{Models from Althaus \& Benvenuto (1997).}
\tablenotetext{c}{Models from Hansen \& Phinney (1998).}
\tablenotetext{d}{Models from Driebe et al. (1999).}
\tablenotetext{e}{Values for each star, assuming two identical stars.}
\end{deluxetable}

\end{document}